# Cardiac Segmentation from LGE MRI Using Deep Neural Network Incorporating Shape and Spatial Priors

Qian Yue, Xinzhe Luo, Qing Ye, Lingchao Xu, and Xiahai Zhuang[✉]

School of Data Science, Fudan University, 200433 Shanghai, Shanghai China
`zxh@fudan.edu.cn`

**Abstract.** Cardiac segmentation from late gadolinium enhancement MRI is an important task in clinics to identify and evaluate the infarction of myocardium. The automatic segmentation is however still challenging, due to the heterogeneous intensity distributions and indistinct boundaries in the images. In this paper, we propose a new method, based on deep neural networks (DNN), for fully automatic segmentation. The proposed network, referred to as SRSCN, comprises a shape reconstruction neural network (SRNN) and a spatial constraint network (SCN). SRNN aims to maintain a realistic shape of the resulting segmentation. It can be pre-trained by a set of label images, and then be embedded into a unified loss function as a regularization term. Hence, no manually designed feature is needed. Furthermore, SCN incorporates the spatial information of the 2D slices. It is formulated and trained with the segmentation network via the multi-task learning strategy. We evaluated the proposed method using 45 patients and compared with two state-of-the-art regularization schemes, i.e., the anatomically constraint neural network and the adversarial neural network. The results show that the proposed SRSCN outperformed the conventional schemes, and obtained a Dice score of 0.758±.227 for myocardial segmentation, which compares with 0.757±.083 from the inter-observer variations.

## 1 Introduction

Analysis of myocardial (Myo) viability is crucial to better understand the physiological and pathological processes for patients suffering from myocardial infarction (MI). Late gadolinium enhancement (LGE) MRI is a valuable tool for MI assessment, because it can visualize the important pathological information. For quantitative assessment, segmentation of the myocardium is a prerequisite.

Manual segmentation can be time-consuming and suffer from inter-observer variations, thus automating this process is desirable in the clinic. Rajchl *et al.* (2014) proposed to segment the myocardium indirectly using a multi-region approach [1]. Many automatic methods use cine MRI as prior knowledge, and the image registration techniques are applied for more accurate segmentations [2]. These methods generally require an accurate registration between the cine MRI and LGE MRI. However, this registration can also be challenging, considering the intra-image misalignments as well as inter-image misregistration. Therefore, manual interaction is commonly used. Liu *et al.* employed the multi-component Gaussian mixture model to automatically segment the myocardium from a single LGE MRI sequence [3]. The coupled level set is employed as a spatial constraint, which can be iteratively adapted according to the image characteristics.

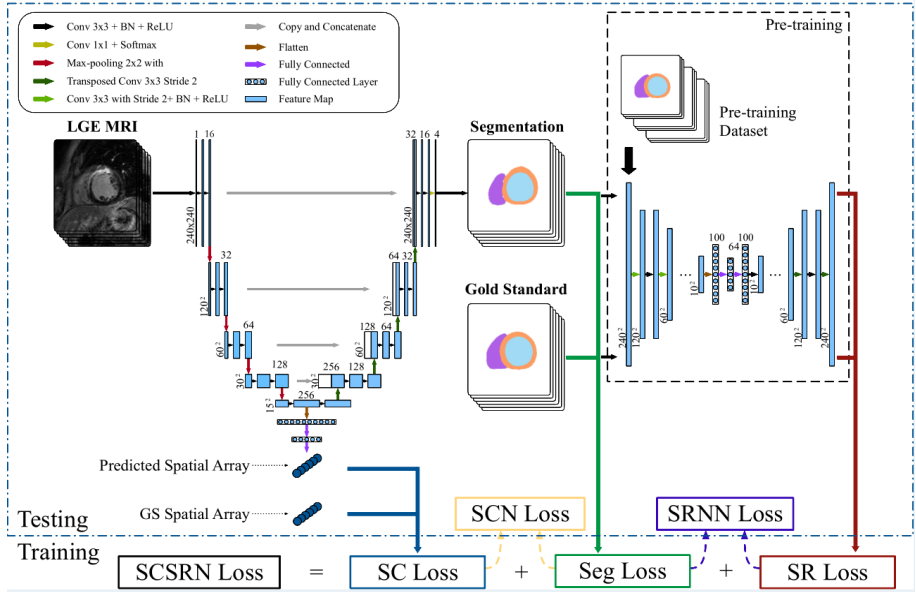

**Fig. 1.** Overall structure of SRSCN, whose loss comes from three parts: the segmentation loss is specially design as a function of cross entropy and Dice, the spatial constraint (SC) loss to assist segmentation, and the shape reconstruction (SR) loss for shape regularization.

Fully automated segmentation of LGE MRI is challenging due to the heterogeneous intensity distributions of images and the large shape variation of the heart. Furthermore, the annotated data are meanwhile limited; thus, the attempt to solve this problem automatically is still rarely reported. In the field of medical imaging, anatomical priors can be essential in assisting the segmentation task in the deep neural network (DNN)-based algorithms. Therefore, in this work we propose an enhanced DNN model with shape reconstruction (SR) and spatial constraint (SC) to tackle the challenging segmentation task, particularly with a small set of annotated training data. The resulting network is expected to be able to constrain the segmentation to generate results with realistic heart shapes.

We first propose a shape reconstruction neural network (SRNN). SRNN can be pre-trained by anatomical priors such as a set of label images, and it works as a shape constraint to regularize the results. Hence, SRNN can maintain a realistic heart shape of the segmentation result. Furthermore, we propose the spatial constraint network (SCN) to solve the large variation of the 2D slices across different positions of a 3D cardiac MRI. This is because the 2D slices may come from any position, from the apex to the base of the ventricles. The shape and appearance of these slices can vary considerably if they come from different positions. SCN is designed to incorporate this information. By combining the learning task of spatial information with the segmentation problem and formulating them as a two-task-learning problem, one can expect the SCN to significantly improve the general performance of the network, opposed to the separate training for the two tasks. In addition, we investigate two state-of-the-art alternatives for shape regularization, i.e. the anatomical constraint neural network (ACNN) [4] and the generative adversarial network (GAN) [5], though neither of them has been used for this segmentation task, to the best of our knowledge.

## 2    Method

Fig. 1 presents the structure of the proposed network, i.e., SRSCN, which is based on an enhanced U-Net [6]. SRSCN includes two modules to incorporate the prior knowledge, i.e., the SR module and the SC module. The models solely combining U-Net with SR and SC are denoted as SRNN and SCN, respectively.

### 2.1    Architecture of the Segmentation Network

Ronneberger *et al.* proposed U-Net for medical image segmentation, which has two key modules, i.e. the feature extraction and up sampling module [6]. Based on the fully connected network (FCN), it has the advantage of utilizing multi-scale information of the images. U-Net has a symmetric pyramid structure, where an input image is compressed into higher semantic features and then unsampled to its original resolution. The combination of local and contextual information enables a good segmentation of medical images.

In our work, we adopted the Exponential Logarithmic Loss [7] as the loss function to measure the result of the segmentation. This loss function combines cross entropy and Dice score in a balanced fashion to facilitate training, and it takes the label balance into account to accelerate convergence, i.e.,

$$L_{Seg} = \lambda_{Dice}L_{Dice} + \lambda_{Cross}L_{Cross}, \qquad (1)$$

where, $\lambda_{Dice}$ and $\lambda_{Cross}$ are the balancing parameters, respectively for the weighted Dice score term, $L_{Dice} = \boldsymbol{E}[(-\ln(Dice_i))^{\gamma_1}]$, and the weighted cross entropy term, $L_{Cross} = \boldsymbol{E}[w_l(-\ln(p_l(\boldsymbol{x})))^{\gamma_2}]$ with $\boldsymbol{x}$ the pixel position, $i$ the label, and $l$ the ground-truth label at $\boldsymbol{x}$; $\gamma_1$ and $\gamma_2$ are two hyperparameters that control the nonlinearities of the loss functions.

DNN, however, generally requires a large set of annotated data to train the network. With limited training data, the generalization capacity of the network could be impaired. Therefore, constraints from prior knowledge should be included to enhance the performance of the DNN.

## 2.2 SRNN for Prior Knowledge of Shapes

SRNN aims to learn an intermediate representation, from which the original inputs can be reconstructed. Internally, by several down sampling operations, it can compress the information or knowledge of original input into some codes acting as a compact representation of the input image. Through this information compression, features of the inputs are captured and mapped into a high-density space.

Hence, an SRNN model, pre-trained from a set of shape images, is able to function as a constraint to regularize a segmentation result into a desired realistic shape. The architecture of this SRNN is illustrated in Fig. 1, where the SR module (in dark red) is connected, as an extended network to U-Net. During the optimization process, a regularization term produced by SRNN is in charge of constraining segmentation output. The loss function for training SRNN is formulated as follows,

$$L_{SRNN} = L_{Seg} + \lambda_{SR} L_{SR} , \qquad (2)$$

where $\lambda_{SR}$ is the balancing parameter; $L_{SR}$ is the SR module loss and is defined from Frobenius norm,

$$L_{SR} = \sum_{i=1}^{n} \left\| \hat{R}_i - R_i \right\|_F^2 . \qquad (3)$$

Here, $n$ is the number of training samples, $R_i$ indicates the reconstructed gold standard segmentation, and $\hat{R}_i$ denotes the reconstructed segmentation from the SRNN prediction; $\|\cdot\|_F$ is the Frobenius norm of an $m \times n$ matrix, and it is defined as the square root of the sum of the absolute squares of matrix elements.

## 2.3 SCN for Prior Knowledge of Spatial Constraints

The idea of utilizing spatial information comes from the fact that the shapes and appearance of the heart in the basal and apical slices can vary significantly. Therefore, we develop an SC module to include the prediction of the spatial information of each slice. At the same time, the segmentation task cooperates with spatial information prediction task, which forms a multi-task learning problem. Multi-task learning has been shown to be able to significantly improve the performance in contrast to learning each task independently, both empirically [8] and theoretically [9,10]. This is the case not only when a few data per task are available but also when two tasks can intuitively strengthen each other.

As Fig. 1 shows, we propose the SC module (in dark blue), connected to the bottom of the U-Net, to predict the position of an LGE MRI slice. The SC loss is designed to penalize the erroneous prediction of the spatial positions,

$$L_{SC} = \sum_{i=1}^{n} \left\| \hat{P}_i - P_i \right\|_F^2 , \qquad (4)$$

where $P_i$ is the ground truth spatial information of slice $i$, and $\hat{P}_i$ is the prediction. Similarly, the SCN loss is formulated with the weighted loss terms,

$$L_{SCN} = L_{Seg} + \lambda_{SC} L_{SC} . \qquad (5)$$

By incorporating SC, the network can combine two tasks, i.e., the regression of position and the segmentation of images, to form a two-task-learning problem.

### 2.4 The proposed SRSCN

Finally, we combine the SRNN and SCN to obtain the SRSCN, as shown in Fig. 1, whose loss function is then defined as follows,

$$L_{SRSCN} = L_{Seg} + \lambda_{SC} L_{SC} + \lambda_{SR} L_{SR}. \tag{6}$$

These two techniques can strengthen each other and result in better segmentation. The two weights, $\lambda_{SC}$ and $\lambda_{SR}$, balance the regularization effect of these two terms.

### 2.5 Alternative Technology for Shape Constraints

For comparisons, we further investigate the two state-of-the-art networks for shape regularization, i.e., ACNN and GAN.

ACNN takes a series of cardiac label images as the inputs [4]. Through the pretrained auto-encoder network, the shape features are encoded as the compact codes of the network. In contrast to the proposed SRNN using the reconstruction to assist segmentation, ACNN solely uses the codes created by the encoder. Specifically, one can obtain the ACNN by replacing the regularization term in SRNN with the L2-norm between the codes coming from the segmentation result and gold standard.

GAN trains a discriminator to distinguish the authenticity of the inputs [5]. The generator of GAN is responsible for producing more realistic inputs to fool the discriminator. Integrating this idea into the segmentation task, it is quite natural to train a discriminator whose task is to identify gold standard and segmentation results. Our main purpose is to guide the segmentation network, that is U-Net to obtain better segmentation results under this regularization. Specifically, two major modifications have been performed on the U-Net to obtain the GAN-regularized U-Net segmentation. Firstly, these segmentation results to be distinguished and gold standard are fed to GAN as it plays the role of predicting a probability determining whether the current input is gold standard label or not. The Sigmoid cross entropy used for GAN penalizes this discriminator for wrong predictions. Secondly, the cost function includes a regularization term created by GAN, with fixed parameters and an input of gold standard label.

## 3 Experiment

### 3.1 Data, Experimental Setup and Implementation Details

The LGE MRI used in the study were collected from 45 patients, of which 25 patients were randomly selected for training, 5 selected for validation and 15 for testing. Note that one of the 15 test cases failed all the methods, due to the particularly poor image quality. Hence, the statistics of the results reported here exclude this outlier. To augment the training data, we registered the training images to other image spaces using a set of artificially generated rigid, affine and deformable transformations, resulting in 1,350 augmented 3D images and 20,405 2D slices.

**Table 1.** Segmentation performance of SRSCN for cardiac LGE MRI.

| Metrics | Myo (Epi) | LV (Endo) | RV (Endo) |
|---|---|---|---|
| Dice | 0.812±0.105 | 0.915±0.052 | 0.882±0.084 |
| ASD (mm) | 1.480±0.997 | 1.749±1.512 | 1.619±1.748 |
| HD (mm) | 11.04±5.818 | 12.25±6.455 | 18.07±14.17 |

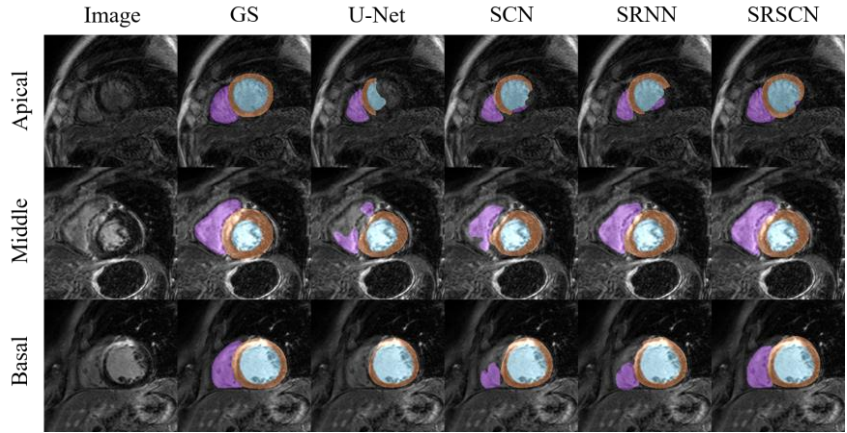

**Fig. 2.** Visualization of three typical slices. Here, GS denotes gold standard segmentation.

We used Dice coefficient, average symmetric surface distance (ASD) and Hausdorff Distance (HD) as metrics for evaluation of segmentation accuracy. ASD measures the average of all the distances from points on the boundary of segmentation (Seg) to the boundary of gold standard (GS),

$$\text{ASD} = \frac{1}{|B_{Seg}|+|B_{GS}|} \times \left( \sum_{x \in B_{Seg}} d(x, B_{GS}) + \sum_{y \in B_{GS}} d(y, B_{Seg}) \right).$$

The HD metric measures how far two subsets of a metric space are from each other, $\text{HD} = \max_{x \in Seg} \min_{y \in GS} \|x - y\|$.

For SRSCN, we used 5e-4 for the weight of SRNN and 1e-6 for SCN as default. Note that it is possible to obtain better performance if an exhaustive search for the optimal value could be employed. The inputs to the networks were 2D slices of size $240 \times 240$ in pixels; the size of mini-batch was 32; the learning rate was 0.001. We trained each model for 30 epochs. GAN was trained for 10 epochs with manual monitoring of convergence, due to the particularly expensive training. The codes and models were implemented using TensorFlow [11], and the optimizer for training was AdamOptimizer [12]. We used one GPU of type GTX 1080ti for training and testing. Each model required 5 to 8 hours to train and the testing of a subject took 2 to 3 seconds.

**Table 2.** Dice scores of the different methods from the study of shape constraints.

| Methods | LV | RV | Myo | Mean |
|---------|-----|-----|-----|------|
| U-Net | 0.816±0.177 | 0.712±0.272 | 0.682±0.200 | 0.737±0.216 |
| SCN | 0.885±0.119 | 0.797±0.170 | 0.773±0.156 | 0.818±0.148 |
| SRNN | 0.910±0.051 | 0.825±0.122 | 0.796±0.115 | 0.844±0.096 |
| SRSCN | 0.915±0.052 | 0.882±0.084 | 0.812±0.105 | 0.870±0.080 |
| ACNN | 0.913±0.044 | 0.835±0.102 | 0.800±0.088 | 0.849±0.078 |
| GAN | 0.885±0.109 | 0.792±0.190 | 0.781±0.154 | 0.819±0.151 |

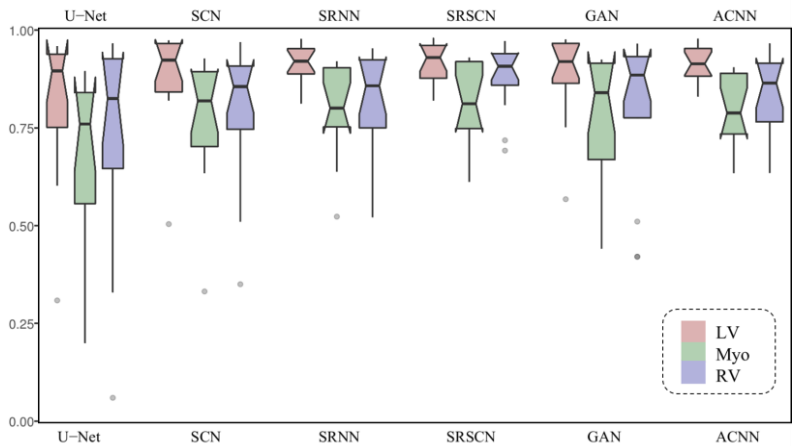

**Fig. 3.** Box plots of the Dice scores of the different methods.

### 3.2 Performance of the Proposed Method

Table 1 presents the statistics of the three metrics of the proposed SRSCN. Dice score for myocardium segmentation reaches 0.812±.105, which compares the inter-observer Dice of 0.757±.083. Note that the mean Dice score drops to 0.758±.227 if the one failure case is included.

### 3.3 Study of Constraints

#### 3.3.1 Ablation study of SRSCN.

The results of the ablation study are presented in Table 2. SCN outperforms U-Net by 8% in terms of generalized Dice score. SRNN further improves Dice performance by 3%. The proposed model, which consists of both of the SR and SC modules achieves more than 13% improvement. Fig 2 visualizes three typical slices, i.e. from apical, middle and basal ventricle, and Fig 3 compares the distributions of Dice scores of different methods. The segmentation improvements are evident in the ablation study.

#### 3.3.2 Comparisons with two state-of-the-art models.

Table 2 and Fig 3 also present the segmentation results from the two state-of-the-art deep-learning-based algorithms, i.e. ACNN [4] and GAN [13]. Compared to ACNN, SRSCN obtains marginally better mean Dice; compared to GAN, it achieves more than 5% improvement. Compared to U-Net without shape regularization, SRSCN has evidently and significantly better Dice scores in all categories ($p<0.01$).

## 4    Conclusion

In this work, we propose the SRSCN for cardiac segmentation of LGE MRI. SRSCN incorporates the shape and spatial priors via the SC and SR modules. SC module is introduced as a spatial constraint for 2D slices and is formulated in the unified loss function as a multi-task-learning problem. SR aims to maintain a realistic shape of the resulting segmentation. We have evaluated the proposed method using 45 patients, and compared it with two state-of-the-art regularization schemes, i.e., ACNN and GAN. The results have demonstrated the effectiveness of the SR and SC regularization terms, and showed the superiority of segmentation performance of the proposed SRSCN over the conventional schemes.